\documentclass[reprint,pra,longbibliography,superscriptaddress,nofootinbib]{revtex4-2}

\usepackage{lipsum}
\usepackage{fancyhdr}
\usepackage{graphicx}
\usepackage[caption=false]{subfig}
\usepackage{braket}
\usepackage{booktabs}
\usepackage{multirow}
\usepackage{bm}
\usepackage{mathtools}
\usepackage{amsmath, esint}
\usepackage{amssymb}
\usepackage{hyperref}
\usepackage{esvect}
\usepackage{float}
\usepackage{placeins}
\usepackage{verbatim}
\usepackage{epstopdf}
\usepackage[normalem]{ulem}
\usepackage{mwe}
\usepackage{array}
\usepackage{soul}
\usepackage{wasysym}

\hypersetup{
    colorlinks=true,
    linkcolor=blue,
    filecolor=magenta,      
    urlcolor=cyan,
    citecolor={blue},
    }

\allowdisplaybreaks

\makeatletter

\usepackage{comment}
\let\wfs@comment@comment\comment
\let\comment\@undefined

\usepackage[final]{changes}
\@namedef{Changes@AuthorColor}{red}
\colorlet{Changes@Color}{red}

\let\wfs@changes@comment\comment
\let\comment\@undefined

\newcommand\comment{%
    \ifthenelse{\equal{\@currenvir}{comment}}
    {\wfs@comment@comment}
    {\wfs@changes@comment}%
}

\makeatother

\begin{document}

\title{Inverse designed photonic crystal waveguides for pulsed operation: \\
dispersion, losses and 
controlled light-matter interactions}

\author{Dominic Thompson}
\email{19djt@queensu.ca}
\affiliation{Centre for Nanophotonics, Department of Physics, Engineering Physics and Astronomy, Queen's University, Kingston, Ontario, Canada, K7L 3N6}

\author{Stephen Hughes}
\affiliation{Centre for Nanophotonics, Department of Physics, Engineering Physics and Astronomy, Queen's University, Kingston, Ontario, Canada, K7L 3N6}

\author{Nir Rotenberg}
\affiliation{Centre for Nanophotonics, Department of Physics, Engineering Physics and Astronomy, Queen's University, Kingston, Ontario, Canada, K7L 3N6}

\begin{abstract}
Photonic crystal waveguides (PCWs) are a powerful platform for optical technologies because
they can spatially confine light on sub-wavelength scales and manipulate the group velocity of propagation modes, both of which enhance light-matter interactions. Many applications in photonics require a large bandwidth of low-loss and constant-velocity slow light, a significant challenge for previous dispersion and Bloch mode engineering techniques. By combining inverse design with an efficient mode solver and physics based formulas, we reduce the computational time of PCW designs by more than $100$ times, allowing for the realization of PCWs with up to an order of magnitude increase in bandwidth and up to $4$ times decrease in loss. We then explore the trade-offs between bandwidth, disorder-induce loss, group index, and dispersion. As examples, we apply this approach to two active and practical areas of research for PCWs design:  broadband, position-tolerant Purcell enhancement, and compact phase shifters for optical communications. Our results significantly improve state-of-the-art PCW designs and provide a general method to optimize PCWs integrated technologies.
\end{abstract}

\maketitle

\section{Introduction}

Photonic crystal waveguides (PCWs) shape light confinement, controlling its flow, spatial structure, and properties at sub-wavelength dimensions \cite{Ohtaka1979, Yablonovitch1987, John1987, Notomi2001}. These properties allow PCWs to enhance nanoscale light-matter interactions, for example, to enhance optical nonlinearities for all-optical signal processing \cite{Krauss2007, Baba2008, Krauss2008, Soljai2004}, sensing applications \cite{Lai2011, Zheng2025}, and efficiently interface with quantum emitters, providing a platform for light-based quantum circuits \cite{Hughes2004,MangaRao2007, PhysRevLett.99.193901,RevModPhys.87.347,PhysRevX.2.011014}.

\begin{figure*}[!tph]
    \centering
    \includegraphics[width=\textwidth]{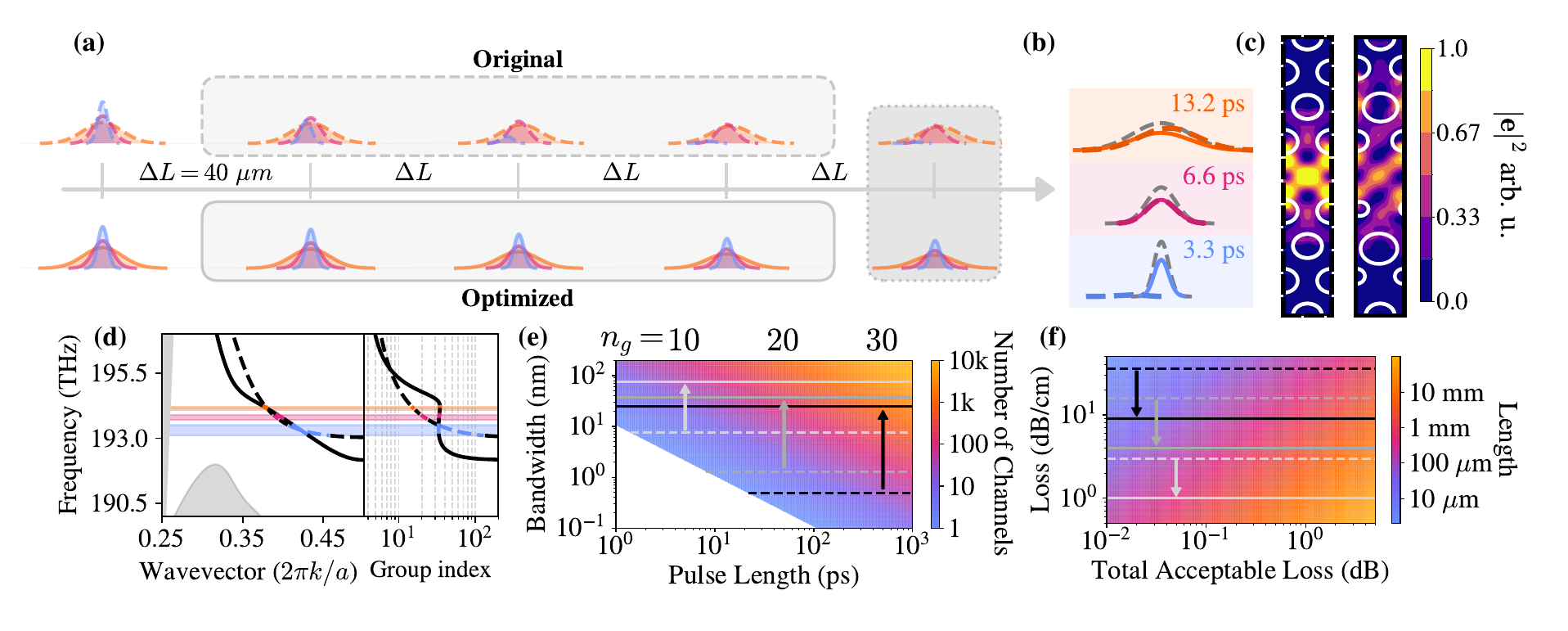}
    \caption{ Inverse design of large-bandwidth, slow-light PCWs.  (a) Calculated pulse profile of Gaussian pulses with different central frequencies and pulse-widths (13.2 ps in orange, 6.6 ps in magenta, and 3.3 ps in blue) travelling down both the original W1 waveguide (top) and an optimized waveguide (bottom). (b) Zoom in on the pulse profile at the end of the PCW, comparing the initial profile (dashed grey curve) with those of the unoptimized (dashed colored curve) and optimized (solid colored curve) designs.  (c) Example intensity distribution of the electric field Bloch mode at 193.5 THz of a unit cell of the original (left) and optimized (right) PCWs in the center of the slab.
    (d) The corresponding dispersion relations and group index for the two designs. Dashed lines correspond to the original design and solid lines to the optimized design. The orange, magenta, and blue regions correspond to the pulses shown in (a), the gray region to the left is the light line, and the lower gray region is the edge of the band gap for the optimized structure. (e) The number of channels supported by different design bandwidths for different pulse widths. Each channel width is assumed to be three times the transform-limited frequency bandwidth of the Gaussian pulse. Bandwidths for the original design (dashed lines) and best-case improvements (solid lines) are shown for different $n_g$'s (10 in light gray, 20 in dark gray, and 30 in black). (f) The total device length that can be made given different waveguide losses and total acceptable loss. Best-case improvements to the original device are shown [dashed to solid curves, as in (f)].}
    \label{fig:fig1}
\end{figure*}

Many photonic technologies use short pulses of light, with typical pulse durations ranging from nanoseconds down to 10's of femtoseconds, and differing frequencies may be used at once to increase the device capacity or processing power. For a typical PCW, these can pose a significant challenge, as illustrated in Fig.~\ref{fig:fig1}(a). Here we observe the calculated profile of three Gaussian pulses [lengths of 3.3 (blue), 6.6 (megenta), and 13.2 ps (orange)] as they propagate through a typical 120~$\mu$m W1 PCW [top panel in Fig.~\ref{fig:fig1}(a); W1 geometry and mode profile shown in the left panel of Fig.~\ref{fig:fig1}(c)]. Both the shape and relative temporal position of each pulse change as it propagates, as is most evident in the zoom in on the final positions [Fig.~\ref{fig:fig1}(b)] where we show both the initial (dashed grey) and final pulse (dashed color) profiles. In particular, the length of the shortest pulse has increased from 3.3 ps to 5.6 ps, and its amplitude has {\it almost completely attenuated}. 

The pulse reshaping occurs due to two main effects. First, in the typical PCW, different frequency components propagate with different velocities, because of dispersion, which causes the pulse to pull apart and broaden. This can be seen in the dispersion relation of the waveguide, shown in Fig.~\ref{fig:fig1}(d) in the dashed curve, along with the group index $n_g$ of the mode. In particular, note that the shortest pulse, whose bandwidth is denoted by the blue shaded region, spans a range of $n_g$ from 33 to 130. Second, PCWs can be very lossy because of disorder-induced scattering, particularly for frequencies near the band-edge where light slows down and light-matter interactions are enhanced \cite{Gerace2004, Hughes2005, Kuramochi2005, PhysRevB.80.195305}, as is the case for the shorter pulse. 

Disorder-induced losses are inevitable and arise due to the interaction of the 
propagating light field with the edges of the holes 
[cf.~Fig.~\ref{fig:fig1}(b)] and scale with $n_g^2$ \cite{PhysRevLett.102.253903, Mann2015}, as discussed further below. These can significantly limit the practical applications of PCW-based devices, for example in the number of distinct photonic channels that they can support or in their lengths, shown by dashed lines in Figs.~\ref{fig:fig1}(e) and (f), respectively. 

Dispersion engineering of PCWs has focused on overcoming these effects by increasing the bandwidth across which they support slow light and pulling this region away from the band edge \cite{Li2008, Frandsen2006}. Much of this work has involved a
typical {\it brute force approach} where the PCW geometry (e.g., hole sizes or positions) 
is systematically varied, though recent optimization methods such as particle swarm optimization \cite{Hirotani2021}, deep learning approaches \cite{Yan2024, Chen2023}, and gradient-based inverse design \cite{Thompson2026, Nussbaum2021, PhysRevA.106.033514, PhysRevResearch.6.L022065} have begun to partly address this issue. 

In this work, we use the gradient information provided by an efficient mode-solving method, Guided Mode Expansion (GME) \cite{PhysRevB.73.235114, Zanotti2024, Minkov2020}, in conjunction with 
physics-based formulas and powerful gradient-based inverse design methods \cite{Zhang2021} to perform multi-dimensional optimization of PCWs rapidly and efficiently. This enables us to solve for 102 million modes across 25,500 separate optimizations targeting different bandwidths, group indexes, and losses while maintaining single-mode operation, for telecom-wavelength PCWs, namely
near 1.5 microns (corresponding frequencies of around
200 THz).

Our calculations reveal an inherent trade-off between different parameters, which can be optimized for specific applications. As shown by an exemplary optimized waveguide (Fig.~\ref{fig:fig1}(c), right panel), we can find large-bandwidth slow-light modes [here, $n_g\approx 30$; solid lines in Fig.~\ref{fig:fig1}(d)] that drastically increase the bandwidth by almost 2-orders of magnitude (Fig.~\ref{fig:fig1}(e), solid lines) with a commensurate increase to the number of supported optical channels. Similarly, the losses of the optimized PCW decrease by about an order-of-magnitude across the $n_g=30$ region, as shown in Fig.~\ref{fig:fig1}(f) (solid lines). Altogether, the drastic improvement in PCW performance can be observed in the pulse propagation, which now occurs with minimal reshaping [bottom panel of Fig.~\ref{fig:fig1}(a) and solid colored curves in Fig.~\ref{fig:fig1}(b)].  Below, we describe our optimization approach and systematically analyze the results before applying them to two exemplary applications: enhancing quantum light-matter interactions and a short PCW phase-shifter for a Mach-Zehnder Modulator (MZM). Our results help unveil the limitations and power of PCW engineering, providing a route to low-loss, high-bandwidth designs for enhanced nanoscale light-matter interactions.

The rest of our paper is organized as follows:
In Sec.~\ref{sec:background}, we summarize the basic background theory needed to understand PCWs and define how the backscattering loss (and other important design metrics) can be efficiently computed. In Sec.~\ref{sec:SL}, we define practical constraints of the design problem, and show how we set up the inverse design algorithm to optimize efficiently. In Sec.~\ref{sec:results}, the results from three different classes of optimizations are shown: (i) general PCW design, (ii) PCWs for certain on-chip quantum optics applications, and (iii) PCWs for communications. Finally, in Sec.~\ref{sec:conclusion}, we summarize the key results and findings.

\section{Background Theory}\label{sec:background}

In this section, we briefly introduce the basic theoretical concepts and methods used with our inverse design techniques, including the Bloch modes, disorder-induces losses, 
and important dispersion properties,
which all tie in analytically to the concept of using and optimizing the modes
and PWC band structure.

\subsection{Photonic crystal waveguide Bloch modes}

\begin{figure}[ht]
  \centering
  \includegraphics[width=.9\columnwidth]{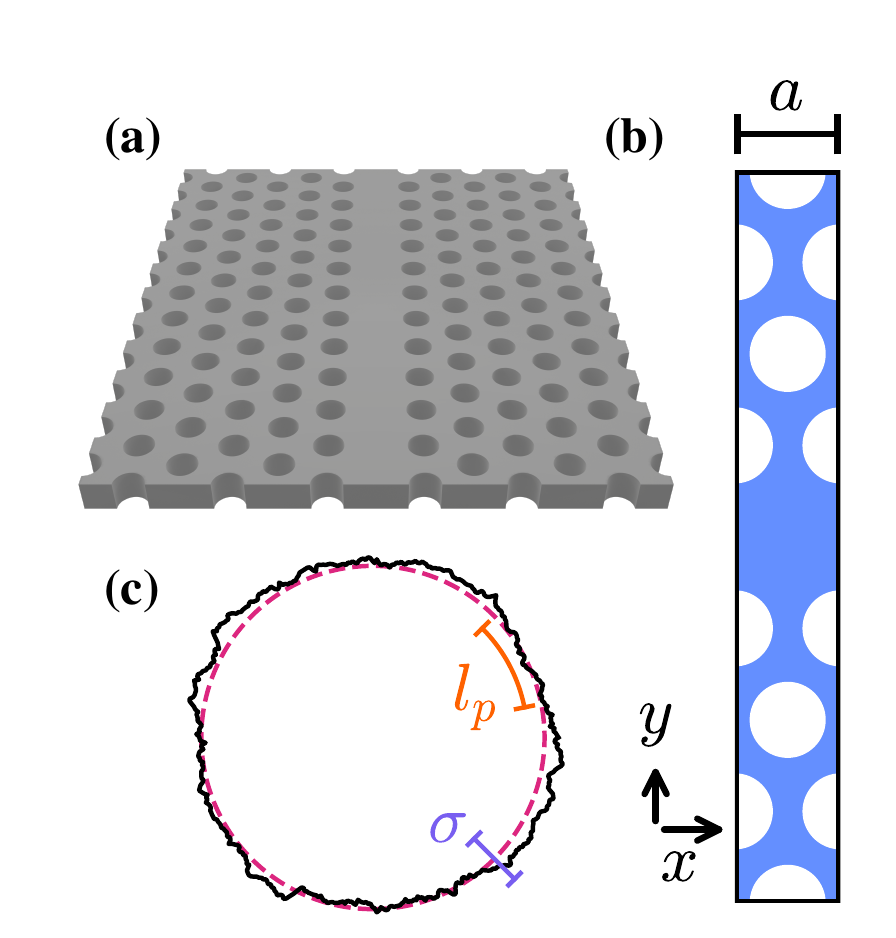}
  \caption{Geometric structure of PCW slab. (a) View of a section of the full 3D PCW slab. (b) Unit cell of a W1 PCW with the lattice constant $a$. The direction of propagation is $x$. (c) Schematic of the fabrication roughness.}
  \label{fig:struct}
\end{figure}

Photonic crystal waveguide slabs are a 2D lattice of holes etched into a slab of finite height, depicted in Fig.~\ref{fig:struct}(a), shown with {\it air cladding}. Along the direction of propagation, PCWs are periodic, as seen in Fig.~\ref{fig:struct}(b), and therefore, the magnetic and electric fields can be written as  Bloch modes:
\begin{align}
    \bm H_k(\bm r)={\bf u}_k(\bm r)e^{ikx/2\pi}\\
    \bm E_k(\bm r)={\bf e}_k(\bm r)e^{ikx/2\pi},
\end{align}
where $k\in [-a^{-1},a^{-1}]$ is the wavevector along the direction of propagation and $a$ is the lattice constant depicted in Fig.~\ref{fig:struct}(b). To compute these Bloch modes, we use GME to efficiently solve the magnetic field eigenvalue problem \cite{PhysRevB.73.235114}.
which in turn gives us the electric fields.

To be more specific, 
for lossless and frequency-independent media, the magnetic field modes are solutions to 
\begin{equation}
    \bm\nabla\times\left(\frac{1}{\epsilon(\bm r)}\bm\nabla\times\bm H_{k}(\bm r)\right)=\left(\frac{\omega}{c}\right)^2\bm H_{k}(\bm r).
\end{equation}
Once the magnetic field modes are solved for, they are used to compute the electric field modes,
\begin{equation}
    \bm E_{k}(\bm r)=\frac{i}{\omega_k\epsilon_0\epsilon(\bm r)}\bm \nabla\times\bm H_{k}(\bm r).
\end{equation}
This gives both the mode profile and dispersion relation, [cf.~Fig.~\ref{fig:fig1}(c) and (d), respectively]. We choose to normalize the (electric field) Bloch mode such that 
\begin{equation}\label{eq:Enorm}
    \int_{\text{cell}}\epsilon(\bm r)|{\bf e}_{k}(\bm r)|^2d\bm r=1,
\end{equation} 
where the integral is carried out over one unit cell (along the direction of propagation), as shown in Fig.~\ref{fig:struct}(b).

We also highlight that GME is remarkably efficient and accurate for
high-index semiconductor PCW slabs (where
$\epsilon\approx 12$ with air holes $\epsilon=1$) as shown in
\cite{PhysRevB.73.235114,Vasco2018,PhysRevB.95.224202}, and has been used as input
to Bloch mode expansion techniques to explain various experiments, including Anderson localization modes~\cite{Crane2017}.

\subsection{Photonic crystal waveguide dispersion properties}

From the dispersion relation, the group index can be computed (or also from the modes directly). The group index $n_g$ defines the ratio between the speed of light in a vacuum, $c$, and the group velocity of the envelope of light within the PCW, $n_g=\frac{c}{v_g}$. We calculate $n_g$ from the band structure according to,
\begin{equation}\label{eq:ng}
    n_{g,k}=c\left(\frac{d\omega_k}{dk}\right)^{-1},
\end{equation}
with larger $n_g$ values corresponding to slow light propagation. 

Unlike conventional waveguides, PCWs have very small regions of consistent group index, as seen in Fig.~\ref{fig:fig1}(d); here, we define the usable bandwidth as the range of wavelengths $\Delta \lambda$ where the group index stays within a certain range $\Delta n_g$ of a {\it target} value, $n_g^t$. With these, we can define the typical figure of merit for bandwidth, the {\it normalized delay bandwidth product} ($\mathrm{NDBP}$) as 
\begin{equation}
    \mathrm{NDBP}=\frac{\Delta\lambda}{\lambda_0}n_g^t \, ,
\end{equation}
where $\lambda_0$ nm is the central wavelength. The relative {\it flatness} of the band, or speed of light, across this bandwidth can be quantified by the dispersion, which is calculated according to
\begin{equation}\label{eq:beta}
    \beta_2=\left(\frac{d^2\omega_k}{dk^2}\right)^{-1}.
\end{equation}

With respect to dispersion engineering, our goal is to maximize the $\mathrm{NDBP}$ for large $n_g$'s, while maintaining a relatively small $\beta_2$.

\subsection{Disorder-induced losses}

Large group index is a key feature of PCWs that makes them promising for many applications, as seen in Sec.~\ref{sec:QDs} and \ref{sec:MZMs}; however,  large group index also leads to large backscattering losses, the dominant form of loss in slow light PCWs~\cite{Hughes2005, Kuramochi2005, PhysRevB.80.195305}. The expected backscatter loss arises from the field scattering off imperfections on the hole surfaces \cite{Gerace2004, Hughes2005}. The disorder along the hole edge is characterized by a correlation function between different points on the edges of the holes, $\tilde \phi$ and $ \tilde\phi'$. Previous work has shown 
that the following disorder profile is a good representation of the real fabricated samples~\cite{PhysRevB.80.195305, Kuramochi2005} 
\begin{equation}
        \langle\Delta R(\tilde\phi)\Delta R(\tilde\phi')\rangle = \sigma^2\exp\left(\frac{-R_j|\tilde\phi-\tilde\phi'|}{l_p}\right)  \delta_{j,j'} ,
\end{equation}
 to approximate natural roughness seen from fabrication, where $j$ indexes the holes and $R_j$ is the radius of each hole. 
 
 The statistical parameters of the roughness are $\sigma$, the standard deviation of the roughness, and $l_p$, the correlation length of the roughness, depicted in Fig.~\ref{fig:struct}(c). 
The change in permittivity can be derived from the change in radius through 
\begin{align}
    \Delta\epsilon(\bm r)=&(\epsilon_2-\epsilon_1)\Theta\left(\frac{h}{2}-z\right)\nonumber\\&\times\sum_j\Delta R(\tilde \phi) \delta\left(R_j-|\bm \rho -\bm \rho_j|\right),
\end{align}
where $\epsilon_1$ and $\epsilon_2$ are the permittivity of the slab and cladding materials, respectively, $\bm \rho_j$ is the in slab vector that defines the center of the hole, $\bm \rho$ is the in-slab position vector, and $\Theta$ restricts the perturbation to within the slab height $h$. 

Since the roughness is much smaller than the full device, it has been shown that the backscattering loss can be computed efficiently using perturbation theory combined with a Green's function scattering theory \cite{Hughes2005, PhysRevLett.102.253903}.
This analytical approach is what makes the problem tractable and extremely efficient with a full 3D model.
The backscatter loss can then be computed for a nominally lossless mode as 
\begin{widetext}
\begin{align}\label{eq:backFull}
    \langle\alpha_{\text{back}}\rangle_k
    &=\sum_{j}\bigg(\frac{a\omega_kn_{g,k}\sigma}{2}\bigg)^2(\epsilon_2-\epsilon_1)^2\int\int_{\text{cell}}d\bm{r}d\bm{r}'
    \Theta\bigg(\frac{h}{2}-|z|\bigg) \Theta\bigg(\frac{h}{2}-|z'|\bigg)\nonumber \\
    & \times \delta(R_j-|\bm{\rho}-\bm{\rho}_j|) [{\bf e}^*_{k}(\bm{r})\cdot{\bm p}^*_{k}(\bm{r})][{\bf e}_{k}(\bm{r}')\cdot{\bm p}_{k}(\bm{r}')]
    \exp\bigg(\frac{-R_j|\tilde\phi-\tilde\phi'|}{l_p}+i2k(x-x')\bigg),
\end{align}
\end{widetext}
where $\omega_k$ is the mode frequency, $x$ is the position along the direction of propagation, the integral is performed over one unit cell, and $\bm p_k(\bm r)$ is the polarization density \cite{PhysRevLett.102.253903, Mann2015}: 
\begin{equation}
    {\bm p}_k(\bm r)=\left[{\bf e}_{k,||}(\bm r)+\epsilon(\bm r)\frac{{\bm d}_{k,\perp}(\bm r)}{\epsilon_1\epsilon_2}\right],
\end{equation}
where $\bm d_k(\bm r)$ is the electric displacement field of the Bloch mode \cite{Hauff2022, Johnson2002}.

Backscatter loss scales with the square of the group index and the strength of the light field around the hole edges, Eq.~\eqref{eq:backFull}. Therefore, to reduce the backscatter loss for a constant group index, the design challenge becomes reducing the Bloch mode overlap with the edges of the holes, an approach often referred to as Bloch mode engineering \cite{Mann:13, Wang:12, OFaolain:10, Thompson2026}.

\section{Practical Optimizations for  slow-light Photonic Crystals Waveguides}\label{sec:SL}
To date, dispersion engineering of PCWs has typically relied on brute force \cite{Hamachi2009, Li2008}, particle swarm \cite{Hirotani2021}, and deep learning \cite{Yan2024, Chen2023} optimization approaches. These, however, rely on systematic sweeps of a few parameters or, in the case of machine learning-based approaches, on large datasets of existing devices.

In contrast, gradient-based inverse design~\cite{Zanotti2024, Minkov2020} allows for efficient, constrained~\cite{Thompson2026} optimization and has recently been used to independently optimize the bandwidth \cite{Vercruysse2020, Nussbaum2021}, Purcell factor \cite{PhysRevA.106.033514}, and loss \cite{Thompson2026} of PCWs. Here, we use inverse design to simultaneously optimize these parameters, providing a first systematic study of the power and limits of this method. 

To do so, we first solve for the Bloch modes $\textbf{e}_k(\textbf{r})$ (Fig~\ref{fig:fig1}(c)) and their associated frequencies $\omega_k$ for a given PCW geometry using GME method implemented through the Python package Legume~\cite{PhysRevB.73.235114, Zanotti2024}. We build up the band diagram for the PCW, shown in Fig~\ref{fig:fig1}(d), by repeating this calculation for different values of $k$. Together, the Bloch modes and band structure allow us to calculate the group index, dispersion, backscatter loss, and bandwidth of the PCW, as discussed in Sec.~\ref{sec:background}. We use these, in turn, to define the constrained cost function for the structure, on which we then optimize.

\subsection{Problem definition and constraints}\label{sec:pdeff}
Although an ideal PCW would typically support a large $n_g$ across a wide $\Delta\lambda$, with minimal $\langle\alpha_{\text{back}}\rangle_k$, there is no single best PCW. This is because, in general, there is a trade-off between these different parameters and so the {\it best} PCW is application dependent. Here, we begin by exploring the general phase-space of PCWs to map out the limits of dispersion and Bloch mode engineering. As we do so, we require that a good PCW have several other features: it should support only a single mode in the spectral region of interest, to preclude inter-mode scattering, and its geometry should only contain features that may be fabricated, placing a lower-limit on the size and proximity of the holes.

We optimize each PCW as follows. First, we choose a target group index, $n_g^t$, and a NDBP, which allows us to define a region of $k$-space, ${\mathcal K}_{\rm tot} = \{k_s,...,k_f\}$, where we require that $n_g =n_g^t\pm\Delta n_g$. This region is a subset of the total range of $k$-space, ${\mathcal K}_{\rm tot} \in [0.25, 0.5]$ over which we calculate the properties of the PCW. We choose $k_s$ and $k_f$ for each NDBP such that they are evenly spaced away from the light line and edge of the Brillouin zone and then define the cost function across this bandwidth as,
\begin{equation}\label{eq:cost}
    f(\bm x;n_g^t,{\mathcal  K}_{\rm tot}) = \sum_{k\in{\mathcal K}_{\rm tot}}\left(n_{g,k}(\bm x)-n_g^t\right)^2,
\end{equation}
where $\bm x$ are the design parameters of the PCW. By minimizing $f$, we ensure that the PCW approaches a design with constant $n_g$ for the targeted NDBP. 

As we optimize $f$, we define several constraints to minimize loss, maintain the photonic mode quality, and ensure the PCW can be fabricated. These are described below.
\begin{enumerate}
\item 
\textbf{Backscatter Loss.} For each PCW, we set a target maximum acceptable loss, normalized to account for the $n_g$ dependence,
\begin{equation}\label{eq:Lt}
    L-L^t \le 0,
\end{equation}
where,
\begin{equation}\label{eq:loss_norm}
    L^t=\langle\alpha_{\text{back}}\rangle^t/n_g^2.
\end{equation}
Note that constraining $L$ instead of $\langle\alpha_{\text{back}}\rangle$ makes our optimization more stable since it decouples the constraint from the cost function.

\item 
 \textbf{Frequency range.} To ensure that the desired mode remains within a frequency range, $[\omega_\mathrm{min},\omega_\mathrm{max}]$, we constrain the optimization such that,
\begin{align}\label{eq:Crange}
    \omega_{k_s}-\omega_\mathrm{max}\le0,\\
    \omega_\mathrm{min}-\omega_{k_f}\le 0.
\end{align}

\item 
 \textbf{Single-mode.} For the band of interest to be single-moded throughout the target bandwidth, it must not overlap in frequency with itself or its neighboring bands. We define a constraint to ensure that the band is monotonically decreasing below the light line for all $k$ in ${\mathcal K}_{\rm tot}$ as
\begin{equation}\label{eq:Cmono}
    \omega_{k_{i+1}}-\omega_{k_i}\le0,
\end{equation} 
where $i$ indexes $\mathcal K_{\rm tot}$. In addition, we define two sets of constraints to prevent the band of interest, $b$, from overlapping with its neighboring bands, $b\pm 1$, defined as
\begin{align}
    \max_{k\in{\mathcal K}_{\rm tot}}\left[\omega_{k_{s},b}-\omega_{k,b+1}\right]\le 0,\\
    \max_{k\in{\mathcal K}_{\rm tot}}\left[\omega_{k,b-1}-\omega_{k_f,b}\right]\le 0.
\end{align} 

\item 
 \textbf{Fabrication.} To ensure that the PCW can be fabricated, we add constraints for minimum hole radius and minimum distance between holes. For each hole $i$, we label these properties as $r_i$ and $d_i$, meaning that we constrain the optimization such that,
\begin{align}
    r_\mathrm{min} - r_i\le 0,\\
    d_\mathrm{min}-d_i\le 0, \label{eq:ConstrFab}
\end{align}
where $r_\mathrm{min}$ and $d_\mathrm{min}$ are set by the fabrication protocols and processes. 
\end{enumerate}

For convenience, we define the vector $\bm g(\bm x)$ to contain all constraints from Eqs.~\eqref{eq:Crange}-\eqref{eq:ConstrFab}. The full optimization problem is then defined as 
\begin{equation} \label{eq:optProb}
\begin{array}{rll}
    &\min_{\bm x} & f(\bm x;n_g^t,{\mathcal K}_{\rm tot}) \\[4pt]
    &\text{subject to} & \bm g(\bm x)\le0 \\[2pt]
    && \max_{k\in{\mathcal K_{\rm tot}}}\ L_k(\bm x) \le L^t \, .
\end{array}
\end{equation}

To solve this constrained optimization problem, we use a gradient descent method available through the open-source library SciPy called 'trust-constr.' This method handles the trade-off between satisfying constraints $\bm g(\bm x)$ and optimizing the cost function $f(\bm x)$ in two steps. At the start of each iteration, a barrier sub-problem is formulated, 
\begin{equation}
    \phi(\bm x) = f(\bm x_t)-\mu_t\sum_i\ln(-g_i(\bm x_t)),
\end{equation}
where $t$ indexes the iteration, each constraint is defined such that $g_i(\bm x)$ should be less than zero, and $\mu_t$ is a scalar that defines the weighting of the barrier term. This problem is solved such that each constraint is violated by no more than a specified barrier tolerance. Once a solution within the barrier tolerance has been found, the step is accepted or rejected depending on whether 
\begin{equation}
    \Psi(\bm x_t)= f(\bm x_t)+\rho\sum_i\max(0,g_i(\bm x_t))^2,
\end{equation}
has decreased significantly from the previous step, where $\rho$ scales the contribution of the constraint violation. We used an initial barrier tolerance of 0.1, $\mu_0=0.1$, and $\rho=1$. 


\section{Results}\label{sec:results}

In this section, we present numerical results from applying the Bloch mode analysis discussed in Sec.~\ref{sec:background} and the inverse design framework laid out in Sec.~\ref{sec:SL} to optimize PCWs. The increased computational efficiency, greater than $100$ times speed up, allows us to provide a detailed exploration of the trade-offs between backscatter loss, group index, NDBP, dispersion, and performance parameters of different application-specific devices. 

We begin in Sec.~\ref{sec:dbTradeoffs} by analyzing the trade-offs for a generic PCW. Then, in Sec.~\ref{sec:QDs} and \ref{sec:MZMs}, we adapt the inverse design method to design PCWs for quantum photonic and optical communication applications, respectively, and analyze the trade-offs therein.

For all results, we use $\sigma=3$ nm and $l_p=40$ nm as the sidewall roughness statistical parameters for loss calculations, which have been shown to accurately model roughness for modern nanofabrication techniques 
\cite{Kuramochi2005, Skorobogatiy:05, PhysRevB.80.195305,  PhysRevLett.102.253903}, and define $\bm x$ to be comprised of the in-plane position and radius of the first three holes on either side of the waveguide region.

\subsection{Dispersion and Bloch Mode Engineering Trade-Offs}\label{sec:dbTradeoffs}
Using the aforementioned constraints in Section~\ref{sec:pdeff}, we explore the limitations and trade-offs of PCW optimization. To do so, we design air-clad InP, $\epsilon=11.997$, PCWs, taking the W1 waveguide with a lattice constant of $a=455$ nm, $R=0.3a$, and a slab thickness of $h=270$ nm as the seed design. We optimize for all combinations of 11 group indices ($n_g^t=10$-$30$ in steps of 2), five NDBPs ($\text{NDBP}=0.2$-$0.5$ in steps of $0.75$), and fourteen normalized target losses ($L^t=0.01$-$0.5$ in steps of $0.01$). 

For the frequency constraints, we choose $\omega_{\rm min}=2\pi\times171.3$ THz and $\omega_{\rm max}=2\pi\times215.5$ THz. This allows for greater freedom for the optimization, and any drift from the desired central frequency of $193.6$ THz can be corrected by changing the lattice constant $a$ after optimization. For each set of group index, NDBP, and loss targeted, we performed 20 separate optimizations with slightly perturbed starting values for $\bm x$ to search the large design space. 

Using this approach, we performed 
{\it 280 optimizations for each of the 55 pairs of group indices and NDBPs, using full 3D calculations}. We allow each to run for a maximum of 500 iterations, solving for the modes at 12 different $k$-points, the size of ${\mathcal K}_{\rm tot}$, in each iteration. Including the computation of all constraints and gradients, this took approximately 400 thousand CPU hours. In comparison, Lumerical FDTD 2025 R2 can solve for the modes of a single $k$-point in roughly 30 CPU minutes, meaning the total computation would take approximately 45 million CPU hours without constraint or gradient computation. That is, GME allows for a greater than 100 times speed up as compared to, for example, Lumerical FDTD,
and even further speed ups compared to finite-element solvers like COMSOL..

\begin{figure}[t]
  \centering
  \includegraphics[width=1\columnwidth]{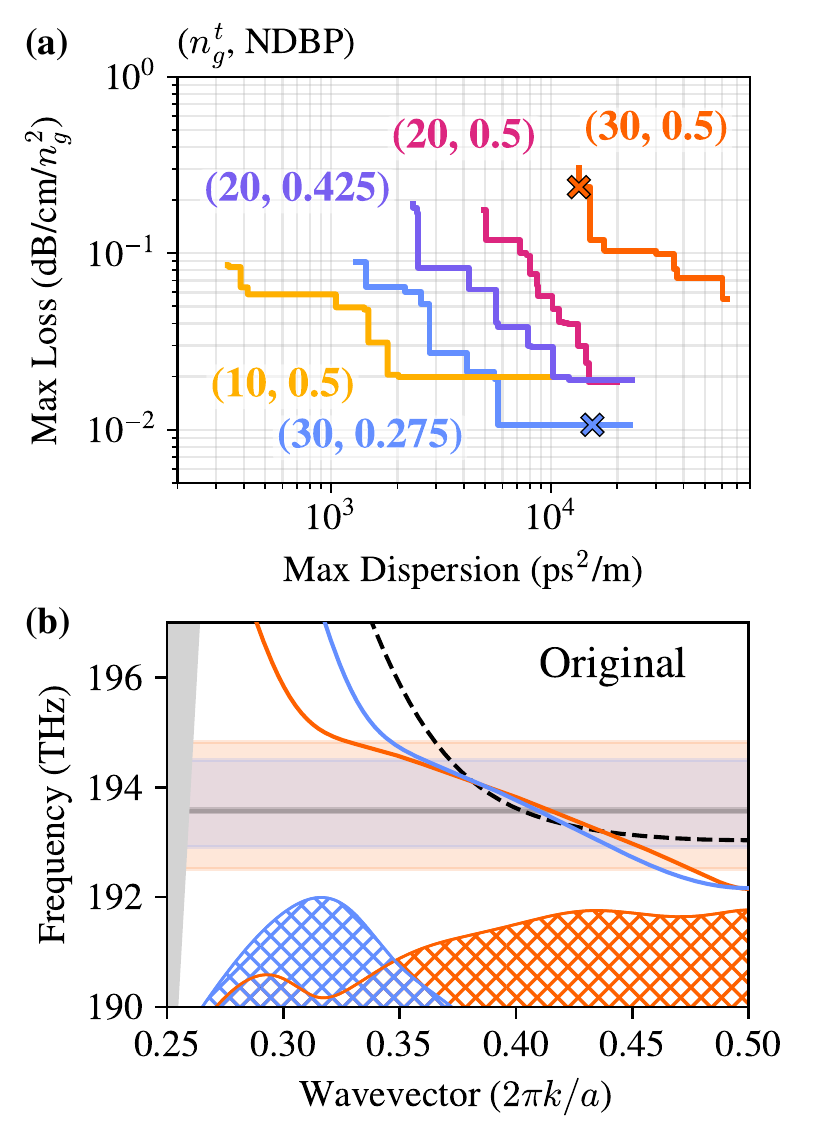}
  \caption{Large-scale optimization of slow-light PCWs. (a) Lowest dispersion and loss designs for a subset of target $n_g^t,\text{NDBP}$ pairs. The orange and blue crosses correspond to the two designs shown in panel (b). (b) Exemplary band diagram for the original (black) and two optimized PCWs designed with $n_g^t=30, \, \text{NDBP}=0.5$ (orange) and $n_g^t=30,\, \text{NDBP}=0.275$ (blue). The shaded horizontal regions show the bandwidths of each design, the gray region is the area above the light line, and the hashed regions are the continuum of modes outside the photonic band gap. The lattice constant of each design has been adjusted to center their bandwidths at $193.6$~THz.}
  \label{fig:fig2}
\end{figure}
\begin{figure}[t]
  \centering
  \includegraphics[width=1\columnwidth]{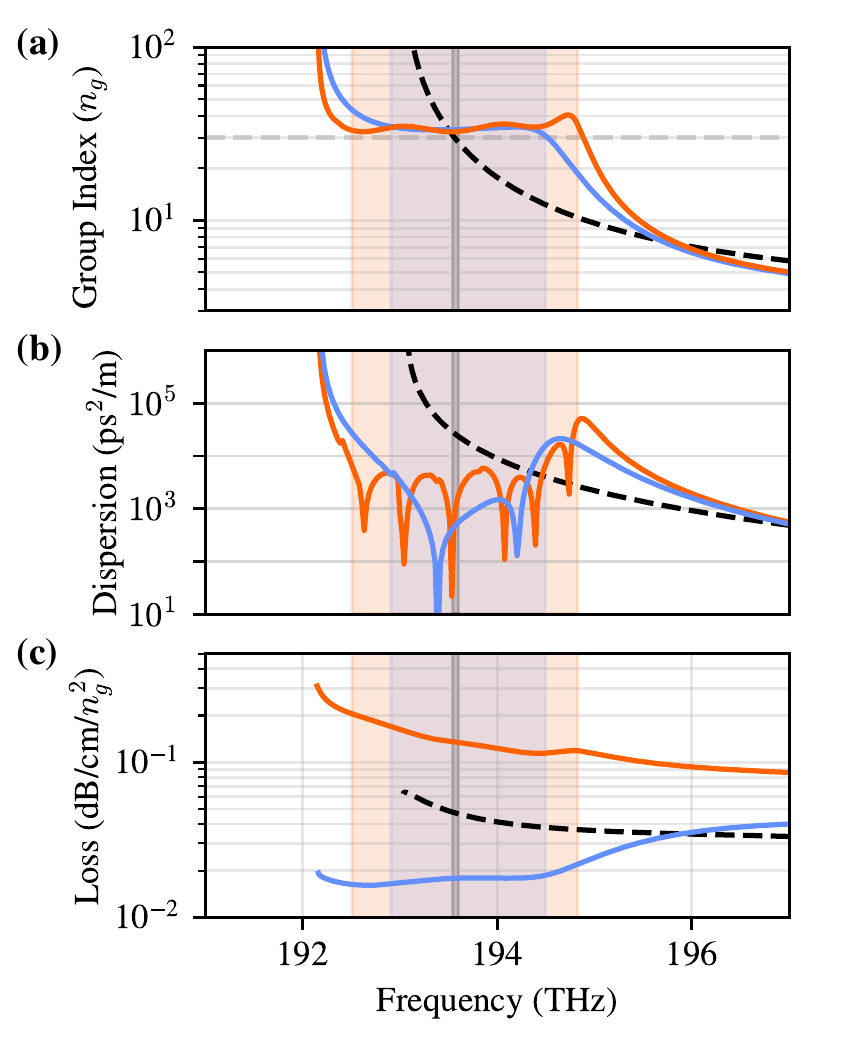}
  \caption{Metrics of optimized slow-light PCWs. The (a) group index, (b) dispersion, and (c) loss of the original W1 (black) and two optimized  PCWs shown in Fig~\ref{fig:fig2} (blue and orange). The shaded regions correspond to their respective bandwidths. The dashed gray line in (a) represents the target group index.}
  \label{fig:fig3}
\end{figure}

Figure~\ref{fig:fig2}(a) shows the results of these design optimizations. Here, each point represents the maximum dispersion versus maximum loss for different optimized PCWs of a given $\left(n_g^t, \mathrm{NDBP}\right)$ pair, where 
\begin{align}
    \text{Max Loss}&=\max_{k\in \mathcal K_{\text{fine}}}L_k,\\
    \text{Max Dispersion}&=\max_{k\in\mathcal K_{\text{fine}}}\beta_{2,k},
\end{align}
where ${\mathcal K}_{\text{fine}}=\{k_s,k_s+\delta k,k_s+2\delta k,...,k_f\}$ and $\delta k=0.00125$. These highlight the trade-off between loss, dispersion, $n_g$, and bandwidth. For instance, if a particular application needed a PCW with $n_g=30$, a maximum loss of $L=0.1\ (\text{dB/cm}/n_g^2)$, and a maximum dispersion of $\beta_2 = 1000\ (\text{ps}^2/\text{m})$, then a PCW with NDBP of $0.275$ could be used (light blue curve), but not a NDBP of $0.5$ (orange curve).

Figure~\ref{fig:fig2}(a) shows that all pairs of group indices and NDBPs have a clear trade-off between dispersion and normalized loss. Within each curve, dispersion may be decreased at a cost of higher losses, and vice versa. Similarly, when either the group index or NDBP is increased, the whole line is shifted to larger dispersions and normalized losses. This is a result of the shrinking design space as stricter constraints are imposed on the loss, or a larger group index or NDBP are desired. These trade-offs are not strictly imposed by any specific theory we are aware of, but are emergent behaviour that exists across all of our optimizations. 

Exemplary dispersion relations for the optimized PCWs that highlight this trade-off are shown in Fig.~\ref{fig:fig2}(b), corresponding to the points marked in Fig.~\ref{fig:fig2}(a). While both designs have a similar group index of 30 and maximum dispersion of $\beta_2\approx1500$ ps$^2$/m, the blue curve supports a smaller NDBP = 0.275 with a small normalized loss of only $0.02$~dB/cm/$n_g^2$. In contrast, the orange curve supports a much larger NDBP = 0.5 at a cost of increased losses of $0.2$~dB/cm/$n_g^2$. Note that, for both designs, we have shifted the lattice constant such that the bandwidths are centred at $193.6$~THz, and that the blue band is used as the optimized example in Fig.~\ref{fig:fig1}(a-d).

While better designs may exist, they are unlikely to yield dramatic improvements to those we find. In fact, our results that we show in Fig.~\ref{fig:fig2}(a) are already drastically better than a conventional W1 PCW. We remind the reader that this improvement can be seen in Figs.~\ref{fig:fig1}~(e) and (f). In the former, we find PCWs with improved bandwidth of 75 nm (from 7.6 nm) for $n_g=10$, 38 nm (from 1.3 nm) for $n_g=20$, and 25 nm (from 0.5 nm) for $n_g=30$. This massively increases the number of channels of the device. For instance, for a pulse length of 10 ps and a group index of 20, the number of channels increases from 1 to 35. Similarly, in Fig.~\ref{fig:fig1}(f) we observe a decreased loss of 1 dB/cm from 3 dB/cm for $n_g=10$, 4 dB/cm from 16 dB/cm for $n_g=20$, and 9 dB/cm from 36 dB/cm for $n_g=30$.

One can clearly visualize the effect of our optimization on the PCW performance in Fig.~\ref{fig:fig3}. Here, we compare the optimized designs for $\left(30, 0.275\right)$ and $\left(30, 0.5\right)$ PCWs [whose dispersion relations are shown in Fig.~\ref{fig:fig2}(b)] to an unoptimized W1 waveguide. First, we observe a large flat bandwidth of relatively constant group index of the optimized designs in comparison to the original waveguide in Fig.~\ref{fig:fig3}(a). Note the slightly larger achieved $n_g$ in both designs, relative to the target $n_g= 30$ (horizontal dashed line), which results from a change of $a$ after optimization to recenter the mode. As NDBP is preserved, this increased $n_g$ results in a small decrease in bandwidth. This could be avoided, in practice, by putting stronger constraints on the minimum and maximum frequency in Eq.~\eqref{eq:Crange}. 

Second, we show the dispersion across the bandwidth in Fig.~\ref{fig:fig3}(b), observing a large decrease across most of the flatband for both optimized designs relative to the initial W1 PCW. We observe that the largest dispersion exists at the edge of the flatband, as the cost function, Eq.~\eqref{eq:cost}, is less dependent on the dispersion (i.e., the derivative of the group index) at its edges. This could be compensated by further constraining the variance of $n_g$ or by considering a slightly larger bandwidth.

Finally, we show the normalized loss as a function of frequency in Fig.~\ref{fig:fig3}(c). For the smaller NDBP = 0.275 design, we achieve lower loss across the entire bandwidth relative to the original PCW design. This is no longer true when NDBP = 0.5, where we find that there is considerably larger loss across the entire flatband, as expected from Fig.~\ref{fig:fig2}(a).

\subsection{Photonic crystal waveguides for applications in quantum photonics}\label{sec:QDs}

Photonic crystal waveguides can act as quantum light-matter interfaces for solid-state quantum emitters \cite{Hauff2022, PhysRevLett.115.153901}. Most often, PCWs are interfaced with semiconductor quantum dots (QDs) to produce high efficiency and indistinguishable single photons for quantum computing and communication applications~\cite{RevModPhys.87.347}, but PCWs could equally be used with other emitters such as defects in diamond \cite{RiedrichMller2014} or silicon \cite{Dobinson2025}. In all cases, PCWs provide two benefits: first, they suppress emission into free space, increasing the coupling efficiency \cite{Wang2011}; 
second, they enhance the emission rate, decreasing the emitter lifetime and drastically reducing the effect of noise on light-matter interactions \cite{Iles_Smith_2017, PhysRevB.98.045309, Albrechtsen2026, Uppu2020}. 

The photon emission enhancement, known as the Purcell factor, can be calculated according to \cite{Hughes2004, MangaRao2007}: 
\begin{equation}\label{eq:PF}
    PF_k(\bm r)=\frac{3\pi c^2n_{g,k}a|{\bf e}_k(\bm r)\cdot \hat {\bf n}|^2}{\omega^2\sqrt{\epsilon}},
\end{equation}
where $c$ is the speed of light in free space, $\bm r$ is the position of the QD, and $\hat {\bf n}$ is a unit vector along the dipole moment of the QD's exciton.

\begin{figure}[t]
  \centering
  \includegraphics[width=1\columnwidth]{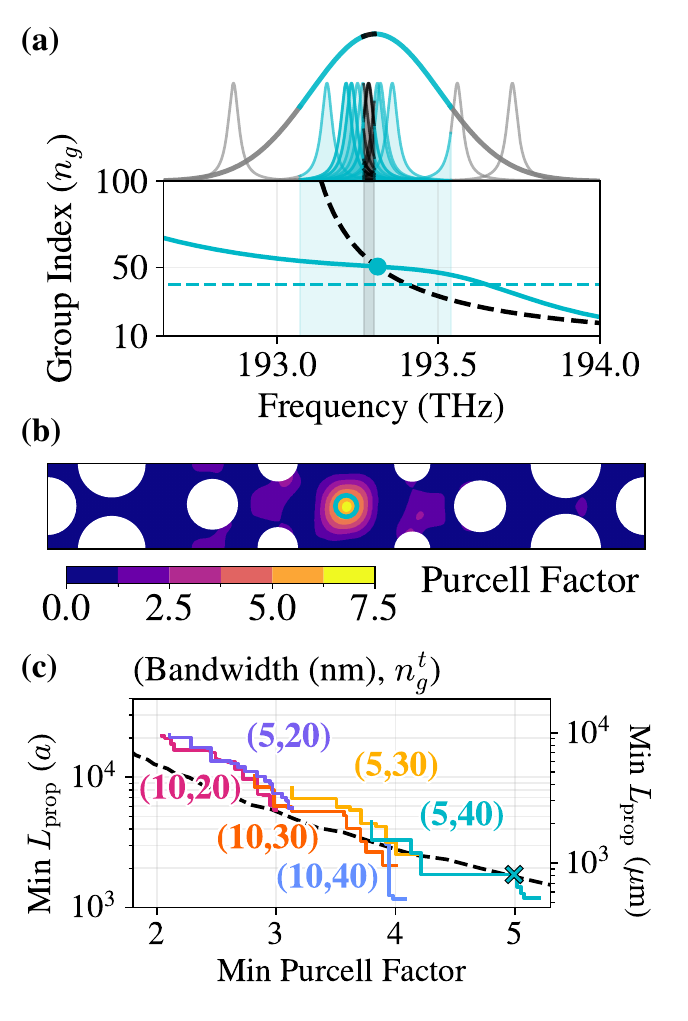}
  \caption{Inverse design for QD-based quantum photonics. (a) The group index as a function of frequency for the original PCW design (black) and an optimized design (cyan). Above the figure, an example emitter inhomogeneous distribution is shown (large, grey gaussian) along with randomly distributed emitter transitions (the small lorentzians). Transitions within the flat bandwidth of each PCW are colored. (b) The spatially dependent Purcell factor in the middle of the slab of the optimized design corresponding to the point shown in (a), for a dipole, $\hat {\bf n}$, oriented orthogonal to the direction of propagation. The circle at the center represents a target 50 nm circle within which we optimize $PF_k$. (c) Achievable propagation length versus minimum Purcell factor for the best optimization results of different bandwidth, $n_g$ pairs, and the original PCW design in black, which has bandwidth ranging from 14 nm near $PF_k=2$ to $0.9$ nm near $PF_k=5$}
  \label{fig:fig4}
\end{figure}

To be useful for applications in quantum photonics, PCWs must provide a large $PF_k$ and low losses across a broad bandwidth. The size of the requisite bandwidth is determined by the inhomogeneous broadening of the emitter, which in turn is determined by factors such as the emitter size or differences in the environment. For example, we sketch a 5 nm broadening in Fig.~\ref{fig:fig4}(a); this is much broader than the typical bandwidth of a W1 PCW, here shown as a grey shaded region. Note, too, that $PF_k$ has a strong position dependence due to the fine structure of the PCW mode. We show this dependence in Fig.~\ref{fig:fig4}(b), highlighting a 50 nm radius circle (cyan curve) at the center where an emitter would experience the largest $PF_k$. Even for emitters with random spatial distributions, such as QDs, a variety of techniques exist for deterministic integration with nanophotonic structures within this accuracy~\cite{Pregnolato2020, Gschrey2013, He2017}, while other emitters such as single defects can be implanted at specified positions \cite{Schrder2017}.

To optimize PCWs for QD-based quantum photonics applications, we follow the same procedure outlined in Sec.~\ref{sec:SL}, adding an additional constraint to force a large Purcell enhancement at the center of the PCW. To do this, we define the {\it normalized} Purcell factor 
(labelled with a tilde)
\begin{equation}
    \widetilde{PF}_k(\bm r)=PF_k(\bm r)/n_{g,k},
\end{equation}
so that the group index is only included in the cost function [see Eq.~\eqref{eq:cost}]. This allows us to set an additional constraint for a target Purcell factor,
\begin{equation}
    \min_{k\in\mathcal K,\bm r\in\Omega}\widetilde{PF}_k(\bm r)\ge\widetilde{PF^t}(\bm r),
\end{equation}
where $\widetilde{PF}^t$ is the target normalized Purcell factor, $\Omega$ is the set of points within 50 nm of the center of the waveguide, shown by the cyan circle in Fig~\ref{fig:fig4}(b), and we take the transition dipole of the emitter, $\hat {\bf n}$, to be oriented in the slab plane orthogonal to the direction of propagation. When this condition is satisfied, any QD within the desired bandwidth and within the region $\Omega$ will experience a normalized Purcell factor of at least $\widetilde{PF}^t$.

We optimize PCWs for 5 nm and 10 nm bandwidths with $n_g^t=20,30$ and $40$, for $L^t=0.01$-$0.05$ in steps of 0.01 and for $\widetilde{PF}^t=0.1$-$0.2$ in steps of 0.025, presenting the results in Fig.~\ref{fig:fig4}(c). Here, for each design we plot the minimum $PF_k\left(\mathbf{r}\right)$ in $\Omega$ versus the calculated loss length in units of $a$ (left axis) and $\mu$m (right axis), comparing to the performance of the original W1 PCW (dashed black curve). Interestingly, the optimization enables us to maintain roughly the same enhancement and loss as the original PCW, but at much greater bandwidths. For the example design shown in Fig~\ref{fig:fig4}(a) [and denoted by a cross in panel (b)], the optimized design has a consistent Purcell factor of greater than 5 across a 3.7 nm bandwidth, as compared to the original design with an equivalent Purcell factor across only a 0.2 nm bandwidth, a more than order-of-magnitude improvement. We note that larger bandwidths, on the order of 10 nm, are possible with a modest decrease in $PF_k\left(\mathbf{r}\right)$ to 4.

\subsection{Photonic crystal waveguides for optical communication applications}\label{sec:MZMs}

Photonic crystal waveguides can also be found in integrated photonic platforms for optical communications and information processing. In such cases, PCWs are increasingly used in compact Mach-Zehnder modulators (MZMs) in silicon platforms as the demand for optical interconnects in data centres grows \cite{Han2022}. These datacom applications require fast, compact MZMs to send data at 10s of GHz across many channels \cite{Shekhar2024}. To achieve the high operating speed required, reverse-biased PN junction MZMs are typically used, but standard rib waveguide designs require large footprints \cite{Reed2010}. The slow light effect in PCWs induces a larger phase shift per unit length than that of rib waveguide MZMs, allowing for smaller components, typically at the cost of small bandwidths and large losses \cite{Kawahara2024, Kawahara2026}. 

\begin{figure}[t]
  \centering
  \includegraphics[width=1\columnwidth]{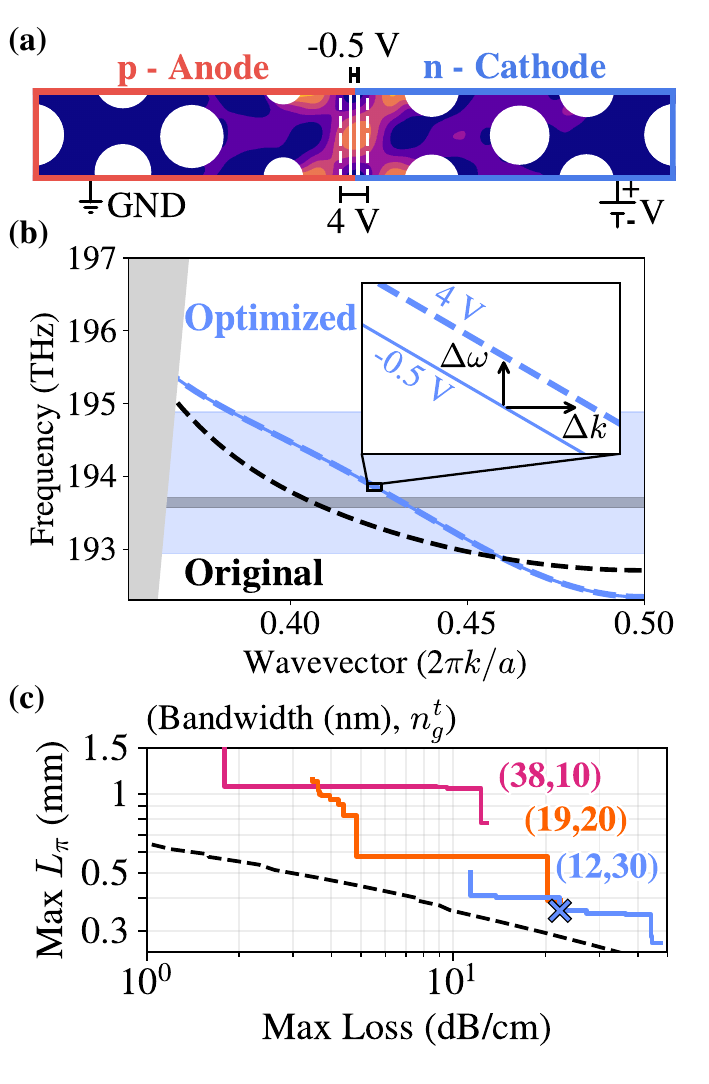}
  \caption{Inverse design for optical communications. (a) The normalized electric field Bloch mode of an optimized PCW with a simple schematic of the p and n-doped regions, electrical connections, and the depletion region at an applied voltage of -0.5V (solid lines) and 4 V (dashed lines). Note the large electric field amplitude contained in these regions. (b) The band diagrams for the original (black, dashed curve) and optimized [solid blue curve; corresponding to the Bloch mode of (a)] designs along with their bandwidths (shaded regions). The inset figure shows the frequency and phase shift of the optimized design caused by the voltage change. (c) Achievable total loss versus $L_\pi$ for optimized PCWs of differing bandwidths and $n_g$'s (colored curves) and for the unoptimized PCW (dashed line). The design used in (a) and (b) is marked with a cross.}
  \label{fig:fig5}
\end{figure}

Reverse-biased PN junction MZMs exploit the plasma dispersion effect in silicon. By modulating the cathode voltage while holding the anode at ground, the width of the depletion region varies, locally changing the refractive index. This causes a phase modulation that accumulates over the length of the MZM. At the output of the MZM, the phase shift causes constructive or destructive interference, allowing for a modulated signal \cite{Reed2010}. The length to induce a $\pi$-phase shift $L_\pi$ depends on the waveguide design, doping concentration, spatial profile of the doping, and the voltage applied to the cathode \cite{Terada2014}. The doping and voltage both have large impacts on the $L_\pi$ of the MZM, but in this work, we fix these parameters and concentrate on the PCW design. We use a linear junction with electron and hole densities of $N_e=N_h= 10^{18}$ cm$^{-3}$ and vary the voltage between $-0.5$ V and $4$ V. This gives a depletion region of $32$ nm for $-0.5$ V and $113$ nm for $4$ V, a change depicted with respect to the unit cell in Fig.~\ref{fig:fig5}(a). 

To model the impact of the carriers and shifting depletion region, we use perturbation theory \cite{Johnson2002, Ramunno2009}. For the mode of a PCW, a small change in the slab permittivity $\Delta\epsilon_{\rm slab}(\bm r)$ corresponds to a change in frequency, 
\begin{equation}
    \frac{\Delta\omega}{\omega_k}=-\frac{1}{2}\int_{\text{cell}}\Delta\epsilon_{\rm slab}(\bm r)|{\bf e}_k(\bm r)|^2dV,
\end{equation}
using the normalization defined in Eq.~\eqref{eq:Enorm}. The change in refractive index of silicon due to the carriers is composed of a real and imaginary part $\Delta \tilde n=\Delta n+i\Delta\kappa$ that can be computed using the experimental results of Ref.~\onlinecite{Nedeljkovic2011}. 

The real component shifts the frequency of the Bloch mode, causing a phase shift as depicted in Fig.~\ref{fig:fig5}(b), and can be computed by 
\begin{equation}
    \Delta\phi_k=\ell n_{g,k}k_0\int_{\text{cell}}\Delta\epsilon_{\rm slab}(\bm r)|{\bf e}_k(\bm r)|^2dV,
\end{equation}
where $\ell$ is the length of the phase shifter \cite{Terada2014}. Similarly, the loss can be computed as 
\begin{equation}\label{eq:dopLoss}
    \alpha_{\text{doping},k}=n_{g,k}k_0\int_{\text{cell}} n\Delta\kappa(\bm r)|{\bf e}_k(\bm r)|^2dV.
\end{equation}
We assume that the doping concentration is zero within the depletion region and equal to the carrier densities outside the depletion region \cite{Mortensen2007}.

We follow the same methodology used previously for the inverse design problem, but impose a slightly different set of constraints. Since the loss from doping depends on $n_g$ [see Eq.~\eqref{eq:dopLoss}] while the backscattering loss depends on $n_g^2$, and we would like to constrain the loss without involving the group index, we define
\begin{equation}
    L_{{\rm tot},k}=\alpha_{\text{doping},k}n_g^t/n_{g,k}+L_k(n_g^{t})^2,
\end{equation}
for the target group indices $n_g^t$ and $L$ as defined in Eq.~\eqref{eq:loss_norm}. This will become accurate as the optimization continues. After optimization, we recalculate the loss with the final group index values, as shown in Fig.~\ref{fig:fig5}(c). For the phase shift, we normalize to the dependence on group index and length so that, 
\begin{equation}
    \theta_k(\bm x)=\frac{\Delta\phi_k}{\ell n_{g,k}}.
\end{equation}

With these definitions, the full optimization problem becomes,
\begin{equation} \label{eq:optProb}
\begin{array}{rll}
    &\min_{\bm x} & f(\bm x;n_g^t,{\mathcal K}_{\rm tot}), \\[4pt]
    &\text{subject to} & \bm g(\bm x)\le0, \\[2pt]
    && \max_{k\in{\mathcal K}_{\rm tot}}\ L_{{\rm tot},k}(\bm x) \le L_{\rm tot}^t,\\[2pt]
    && \min_{k\in{\mathcal K}_{\rm tot}}\ \theta_{k}(\bm x) \ge \theta^t.
\end{array}
\end{equation}

For the PCWs, we use a $220$ nm thick silicon slab $\epsilon=12.1$ with SiO$_2$ cladding $\epsilon=2.085$, with a lattice constant of $a=390$ nm. The maximum and minimum frequencies that we consider are $\omega_{\rm min}=2\pi\times214.7$ THz and $\omega_{\rm max}=2\pi\times 190.1$ THz, and all other constraints within $\bm g(\bm x)$ are the same as defined above. We optimized on $n_g^t=10,20$ and $30$ for one NDBP of $0.25$, using the same procedure as previously stated to determine the components of $\mathcal K$, and five $\theta^t$ values ($\theta^t=0.3$-$0.7$ in steps of 0.1). For $L_{\rm tot}^t$, we use different values that depend on the target group index: for $n^t=10$ we use $L_{\rm tot}^t=15,10,5,3,1.5$, for $n^t=20$ we use $L_{tot}^t=15,10,7.5,5,2.5$, and for $n^t=30$ we use $L_{tot}^t=30,25,20,15,10$. 

We present the result of the optimization in Fig.~\ref{fig:fig5}(b) and (c). Fig.~\ref{fig:fig5}(b) shows an exemplary optimized band structure (blue curve) along with that of the original PCW (dashed black curve). We observe a dramatic increase in the supported bandwidth, which increases from $1$ nm for the original design to $15$ nm. In this case, this order-of-magnitude bandwidth increase comes at the cost of a slightly increased $L_\pi$ and loss, which change from $0.32$ mm to $0.36$ mm and $10.5$ dB/cm to $11.1$ dB/cm, respectively.

There are, however, alternate designs that enable more compact or lower loss devices. This can be seen in Fig.~\ref{fig:fig5}(c), where we present the L$_\pi$ vs. device loss for the different bandwidth-$n_g$ pairs (colored curves) and the original PCW (dashed black curve). Here, we show the loss can be as low as $1.9$ dB/cm, and the device may be as short as $0.26$~mm. There is a trade-off between these two extremes that requires changing bandwidth and group index targets, as seen in Fig.~\ref{fig:fig5}(c).

\section{Conclusions}\label{sec:conclusion}

We have shown how to perform systematic inverse design of PCWs, with example applications in quantum photonics and optical communication. This is done efficiently, reducing computational times by factors of over $100$ compared to standard full 3D FDTD or finite element solver software such as COMSOL. We do this through the use of the GME method, physics-based analytical mode formulas, automatic differentiation, and constrained optimization. The enhanced computational efficiency allows for systematic searching of the design space, which reveals the trade-offs between different figures of merit of the final device. 

Furthermore, this general method for PCW optimization is based on open-source software, allowing it to be easily adapted to most PCW design problems or seed geometries. With this approach, we find designs with up to an order of magnitude increase in bandwidth and up to a $4\times$ reduction in backscatter loss 
[cf.~Fig.~\ref{fig:fig1}(e) and (f)], and are able to map out the design space of these properties, as shown in Fig.~\ref{fig:fig2}(a). 

Turning to exemplary problems in quantum optics and optical information processing, we show broadband and spatially tolerant Purcell factors up to 5.2 for quantum emitter applications [Fig.~\ref{fig:fig4}(c)] and an $L_\pi$ as short as $0.26$ mm for MZM applications [Fig.~\ref{fig:fig5}(c)]. Further, we reveal the trade-offs between Purcell factors and $L_\pi$ with bandwidth and loss. These findings demonstrate the power of using an efficient mode solver with gradient-based, constrained inverse design methods for tackling complicated photonic design problems efficiently.

In contrast to previous approaches, here we perform Bloch mode and dispersion engineering for PCWs simultaneously. Our new framework allows for complete control over the optimized PCW design, allowing PCWs to be designed for specific applications and providing an understanding of design trade-offs. Our work provides a path to the design and fabrication of application-specific PCWs with best-in-class specifications.

\vspace{0.2cm}
\acknowledgements
 This work was supported by the Natural Sciences and Engineering Research Council of Canada (NSERC), the National Research Council of Canada (NRC), the Canadian Foundation for Innovation (CFI), the Ministry of Colleges, Universities, Research Excellence and Security of Ontario, and Queen's University, Canada.

\bibliography{refs}

\end{document}